\RequirePackage{rotating}
\documentclass[iop]{emulateapj}
\usepackage[usenames, dvipsnames]{color}

\usepackage{rotating}
\usepackage[colorlinks=true,linkcolor=black,citecolor=blue,urlcolor=black]{hyperref}%
\usepackage{xcolor}
\usepackage{mhchem}
\defcitealias{matt2015mass}{M15}
\defcitealias{finley2018dipquadoct}{FM18}
\defcitealias{finley2018effect}{Paper I}

\slugcomment{Accepted for publication in The Astrophysical Journal: March 23, 2019}

\shorttitle{Stellar Torque Variability}
\shortauthors{A. Finley, V. See \& S. Matt}

\begin{document}


\title{The Effect of Magnetic Variability on Stellar Angular Momentum Loss II: \\ The Sun, {61 Cygni A}, {$\epsilon$ Eridani}, {$\xi$ Bootis A} \& {$\tau$ Bootis A}}


\author{Adam J. Finley*, Victor See \& Sean P. Matt}
\affil{University of Exeter,
              Devon, Exeter, EX4 4QL, UK}
\email{*af472@exeter.ac.uk}



\begin{abstract}
The magnetic fields of low-mass stars are observed to be variable on decadal timescales, ranging in behaviour from cyclic to stochastic. The changing strength and geometry of the magnetic field should modify the efficiency of angular momentum loss by stellar winds, but this has not been well quantified. In \cite{finley2018effect} we investigated the variability of the Sun, and calculated the time-varying angular momentum loss rate in the solar wind. In this work, we focus on four low-mass stars that have all had their surface magnetic fields mapped for multiple epochs. Using mass loss rates determined from astrospheric Lyman-$\alpha$ absorption, in conjunction with scaling relations from the MHD simulations of \cite{finley2018dipquadoct}, we calculate the torque applied to each star by their magnetised stellar winds. The variability of the braking torque can be significant. For example, the largest torque for $\epsilon$ Eri is twice its decadal averaged value. This variation is comparable to that observed in the solar wind, when sparsely sampled. On average, the torques in our sample range from $0.5-1.5$ times their average value. We compare these results to the torques of \cite{matt2015mass}, which use observed stellar rotation rates to infer the long-time averaged torque on stars. We find that our stellar wind torques are systematically lower than the long-time average values, by a factor of $\sim3-30$. {Stellar wind variability appears unable to resolve this discrepancy, implying that there remain some problems with observed wind parameters, stellar wind models, or the long-term evolution models, which have yet to be understood.}
\end{abstract}



\keywords{magnetohydrodynamics (MHD) - stars: low-mass - stars: stellar winds, outflows - stars: magnetic field- stars: rotation, evolution }


\section{Introduction}
For low-mass stars like the Sun ($M_*\lesssim 1.3M_{\odot}$), magnetic activity is observed to decline with stellar age \citep{hartmann1987rotation, mamajek2008improved}. This is a consequence of the dynamo mechanism, which is responsible for sustaining the stellar magnetic field, and its dependence on rotation and convection \citep{brun2017magnetism}. During the main sequence, angular momentum is removed by magnetised stellar winds. This wind braking increases the observed rotation periods of stars with age \citep{skumanich1972time,bouvier2014angular}. The connections between stellar rotation, magnetic activity and wind braking converges the rotation and activity indices of low-mass stars during the main sequence, such that these quantities appear to follow a mass dependent relationship with age \citep{noyes1984rotation, gilliland1986relation, wolff1997angular, stelzer2001x, pizzolato2003stellar, barnes2010simple, meibom2015spin}. This connection with age is useful in a number of ways. For example, empirical relations can be derived in order to determine the ages of some stars from their rotation or magnetic activity \citep{barnes2003rotational, mamajek2008improved, meibom2009stellar, delorme2011stellar, van2013fast, vidotto2014stellar}.

The observed evolution of rotation also provides a constraint on the torque applied to stars, independently of our understanding of stellar winds. Models for computing the rotational evolution of stars, give us an indication of how stellar wind torques evolve on secular (up to several gigayear) timescales \citep[e.g.][]{gallet2013improved, gallet2015improved}. These torques can then be compared to calculations that are based on observed wind and magnetic properties, in order to test our understanding of stellar magnetism and winds \citep{reville2016age, amard2016rotating}. One caveat, however, is that the torques derived from rotational evolution models are only sensitive to the angular momentum losses of stars averaged over some fraction of the spin-down timescale. For Sun-like main-sequence stars, the rotational evolution torques thus represent a value averaged over $\sim10-100$ Myr. Clearly, any variability of wind and magnetic properties on timescales shorter than this will inhibit a comparison between the long-time torque from rotational evolution models, and those calculated based on observed present-day magnetic and wind properties.

Variability in the magnetic activity of low-mass stars is commonly observed at a range of short timescales, from days to years \citep{baliunas1995chromospheric, hall2007activity, egeland2017mount}. The magnetic fields are driven by the stellar dynamo, whose variability can take many forms, be it in exhibiting a cyclic magnetic field like that of the Sun \citep{saikia2016solar, jeffers2018relation}, magnetic fields with multiple cycles \citep{jeffers2014e} or magnetism with apparently stochastic behaviours \citep{petit2009polarity, morgenthaler2012long}. Such variability appears to occur throughout the main sequence lifetime of low-mass stars. It is therefore interesting to characterise the impact this has on the stellar wind torques.

In order to quantify the impact of magnetic variability on stellar wind braking, we first studied the solar wind in \cite{finley2018effect}, hereafter \citetalias{finley2018effect}, for which we have both in-situ observations of the wind plasma and remote observations of the photospheric magnetic field. In \citetalias{finley2018effect}, available data allowed us to study the variability on timescales from 1 solar rotation ($\sim27$ days) up to a few decades. We quantified how the torque varies on all timescales, and found that the decadal-averaged value was smaller than the rotational evolution torque by a factor of $\sim15$. Although the reason for the discrepancy is still not clear, it could be due to gaps in our understanding of the solar magnetism and wind, variability in the solar torque on timescales much longer than a decade, {issues with the rotational-evolution torques, or a combination thereof.}

In the present paper we examine the influence of observed magnetic variability on the wind braking of four Sun-like stars using semi-analytic relations derived from MHD wind simulations, and compare these values to the long-time average torques derived from modelling the rotational evolution of low-mass stars. In Section 2 we first describe the semi-analytic wind braking formula from \cite{finley2017dipquad, finley2018dipquadoct}, hereafter \citetalias{finley2018dipquadoct}, and the rotational evolution torque prescription from \cite{matt2015mass}, hereafter \citetalias{matt2015mass}. Then in Section 3 we gather stellar properties and magnetic field observations for our four sample stars, each having repeat observations using the Zeeman-Doppler imaging (ZDI) technique. These are {61 Cyg A}, {$\epsilon$ Eri}, {$\xi$ Boo A}, and {$\tau$ Boo A}, three of which also have observed mass loss rates estimated from astrospheric Lymann-$\alpha$ absorption. We also re-examine the Sun, limiting the available data to observations $\sim2$ years apart which is more comparable with the cadence of observations for the other stars. In Section 4 we calculate the angular momentum loss rates using both torque formations, and discuss the results in Section 5.

\section{Angular Momentum Loss Prescriptions}
\subsection{Stellar Wind Torques from Finley $\&$ Matt (2018)}
As in \citetalias{finley2018effect}, we will make use of the semi-analytic formula derived from the MHD simulations of \citetalias{finley2018dipquadoct}. Such formulations are intended for use characterising the braking torques on stars which host convective outer envelopes. {In \citetalias{finley2018effect}, we used a formulation based on the open magnetic flux in the solar wind. Such formulae are independent of the magnetic geometry at the stellar surface \citep{reville2015effect}, however the open magnetic flux cannot be measured for stars other than the Sun. For this work, we instead use a formula based on the observed surface magnetic field instead. Previous formulae, of this kind,} are only valid for single magnetic geometries \citep{matt2008accretion, matt2012magnetic, reville2015effect, pantolmos2017magnetic}, but the magnetic fields of low-mass stars are observed to contain mixed magnetic geometries which vary from star to star \citep[e.g.][]{see2016connection}, and also in time, with geometries evolving in strength with respect to one another \citep[e.g.][for the Sun]{derosa2012solar}.

The \citetalias{finley2018dipquadoct} formulation is simplified, but is capable of approximating the observed behaviour of full MHD simulations without the computational expense. The MHD simulations are performed using axisymmetric magnetic geometries combined with polytropic parker-like wind solutions \citep{parker1958dynamics, pneuman1971gas, keppens1999numerical}, which are relaxed to a steady state. The application of results derived from such simulations to a time-varying problem emulates a sequence of independent steady state solutions. Given that the characteristic timescales for disturbances, caused by the reorganisation of the coronal magnetic field, to propagate through the solution are short with respect to the evolution of the system, this is a valid approximation.

The torque due to a stellar wind is prescribed in terms of the average Alfv\'en radius, $ \langle R_{\text{A}} \rangle$, which acts as an efficiency factor for the stellar wind in extracting angular momentum \citep{weber1967angular, mestel1968magnetic}. The torque, $\tau$, is given by,
\begin{equation}
\tau=\dot M \Omega_{*}R_*^2 \bigg(\frac{\langle R_{\text{A}} \rangle}{R_*}\bigg)^2,
\label{torque}
\end{equation}
where, $\dot M$ is the stellar wind mass loss rate, $\Omega_*$ is the stellar rotation rate (approximated as solid body rotation at the surface), and $R_*$ is the stellar radius. In \citetalias{finley2018dipquadoct}, $\langle R_{\text{A}} \rangle$ is parametrised in terms of the wind magnetisation,
\begin{equation}
\Upsilon=\frac{B_*^2R_*^2}{\dot M v_{\text{esc}}},
\label{up}
\end{equation}
where, the total field strength is evaluated from the first three spherical harmonic components $B_*=|B_{dip}|+|B_{quad}|+|B_{oct}|$, the escape velocity is given by $v_{esc}=\sqrt{2 G M_*/R_*}$, and $M_*$ is the stellar mass. Previous works have shown the reduced efficiency of magnetic braking with increasingly complex magnetic fields \citep{reville2015effect, garraffo2016missing}. {Furthermore, \citetalias{finley2018dipquadoct} examined the behaviour of mixed magnetic geometries and were able to show that higher order modes (e.g. octupole) play a diminishing role in braking stellar rotation, when modelled in conjunction with lower order modes (e.g. dipole, quadrupole).} For mixed geometries, \citetalias{finley2018dipquadoct} showed that the average simulated Alfv\'en radius behaves approximately as a broken power law of the form,
\begin{equation}
  \frac{\langle R_{\text{A}} \rangle}{R_*}=\max\Bigg\{
  \begin{array}{@{}ll@{}}
    K_{\text{dip}}[\mathcal{R}_{\text{dip}}^2\Upsilon]^{m_{\text{dip}}},  \\
    K_{\text{quad}}[(\mathcal{R}_{\text{dip}}+\mathcal{R}_{\text{quad}})^2\Upsilon]^{m_{\text{quad}}}, \\
    K_{\text{oct}}[(\mathcal{R}_{\text{dip}}+\mathcal{R}_{\text{quad}}+\mathcal{R}_{\text{oct}})^2\Upsilon]^{m_{\text{oct}}}.
  \end{array}
  \label{DQO_law}
\end{equation}
This approximates the stellar wind solutions from \cite{finley2018dipquadoct}, for their fit parameters $K_{dip}=1.53$, $K_{quad}=1.70$, $K_{oct}=1.80$, $m_{dip}=0.229$, $m_{quad}=0.134$, and $m_{oct}=0.087$. The magnetic field geometry is input using, $\mathcal{R}_{\text{dip}}$, $\mathcal{R}_{\text{quad}}$, and $\mathcal{R}_{\text{oct}}$, defined as the ratios of the polar strengths for each component over the total field strength, i.e. $\mathcal{R}_{\text{dip}}=|B_{dip}|/B_*$, etc. {We neglect higher order modes than the octupole, as they do not significantly contribute to the torque on the star.}

\begin{table*}[htbp]
\caption{Stellar Parameters}
\label{StellarParameters}
\center
\setlength{\tabcolsep}{1pt}
  \begin{tabular}{c|cccccccc}
  \hline\hline
Star	&	Mass 	&	Radius 	&	Teff	&	$t_{cz}$		&	Rot. Period &Rossby	&	Cyc. Period	&	$\dot M$	\\
Name	&	 $(M_{\odot})$	&	 $(R_{\odot})$	&	(K)	&	 (days)		&	 (days)	&	Number&	 (years)	&	 ($\dot M_{\odot}$)	\\	\hline
Sun	&	1.00	&	1.00	&	5780	&	12.7	&	28 &2.20	&		11	&	1	\\
61 Cyg A	&	0.66	&	0.67	&	4310	&	34.5	&	35.5 &1.03	&	7.3	&	0.5	\\
$\epsilon$ Eri	&	0.86	&	0.74	&	4990	&	24.0	&	11.7	&0.49	&		3.0	&	30	\\
$\xi$ Boo A	&	0.93	&	0.86	&	5410	&	18.3	&	6.4&	0.35		&	7.5$^1$	&	5	\\
$\tau$ Boo A	&	1.34	&	1.42	&	6460	&	1.88 &	3.0	&	1.60		&	0.3	&	$\sim 150^2$	\\
\hline
  \end{tabular}
  \item{$^1$ Fit from this work.}
  \item{$^{2}$ Average mass loss rate from the MHD simulations of \cite{nicholson2016temporal}.}
\end{table*}

\subsection{Rotation Evolution Torques from Matt et al. (2015)}
In this work, we will compare our results to the rotation evolution model of \citetalias{matt2015mass}, which uses the observed distribution of mass versus rotation, at given ages, to find empirical torques that reproduce these observations. To date, no single model (including \citetalias{matt2015mass}) precisely reproduces the observed mass-rotation distributions, but \citetalias{matt2015mass} reproduces the broad dependences of rotation rates on mass and age. The torque in this model has two regimes, either unsaturated where the stellar Rossby number (defined as $Ro=2\pi/(\Omega_*t_{cz})$, where $t_{cz}$ is the convective turnover time) is greater than the saturation value, $Ro_{sat}=0.1R_{o,\odot}$, or saturated where the Rossby number is smaller. All the stars in this paper are in the unsaturated regime. The \citetalias{matt2015mass} torque is given by,
\begin{equation}
\tau = \tau_0\bigg(\frac{t_{cz}}{t_{cz\odot}}\bigg)^p\bigg(\frac{\Omega_*}{\Omega_{\odot}}\bigg)^{p+1} \qquad (Ro_{*} > Ro_{sat}),
\label{unsat}
\end{equation}
\begin{equation}
\tau = \tau_0(10)^p\bigg(\frac{\Omega_*}{\Omega_{\odot}}\bigg) \qquad (Ro_{*} \leq Ro_{sat}),
\label{sat}
\end{equation}
where $p$ is constrained by observations to $\sim2$ \citep{skumanich1972time}, and $\tau_0$ provides the normalisation to the torque based on the stellar mass and radius,
\begin{equation}
\tau_0=6.3\times 10^{30}\text{erg}\bigg(\frac{R_*}{R_{\odot}}\bigg)^{3.1}\bigg(\frac{M_*}{M_{\odot}}\bigg)^{0.5},
\label{norm_torque}
\end{equation}
which is fit empirically from the observed rotation rates of Sun-like stars.

For determining the convective turnover timescales, as in \citetalias{matt2015mass}, we adopt the fit of \cite{cranmer2011testing} to the stellar models of \cite{gunn1998rotation},
\begin{equation}
t_{cz}=314.24\text{exp}\bigg[-\bigg(\frac{T_{eff}}{1952.5\text{K}}\bigg)-\bigg(\frac{T_{eff}}{6250\text{K}}\bigg)^{18}\bigg]+0.002,
\label{tcz}
\end{equation}
where the effective temperature, $T_{eff}$, is the only variable determining $t_{cz}$. \cite{cranmer2011testing} showed that this is a reasonable approximation, which is valid for the temperature range $3300 \leq T_{eff} \leq 7000$ K. Such a monotonic function of $t_{cz}(T_{eff})$ is also supported by other works \citep{landin2010theoretical, barnes2010angular}.

   \begin{figure}
   \centering
    \includegraphics[trim=0cm 0cm 0cm 0cm, width=0.49\textwidth]{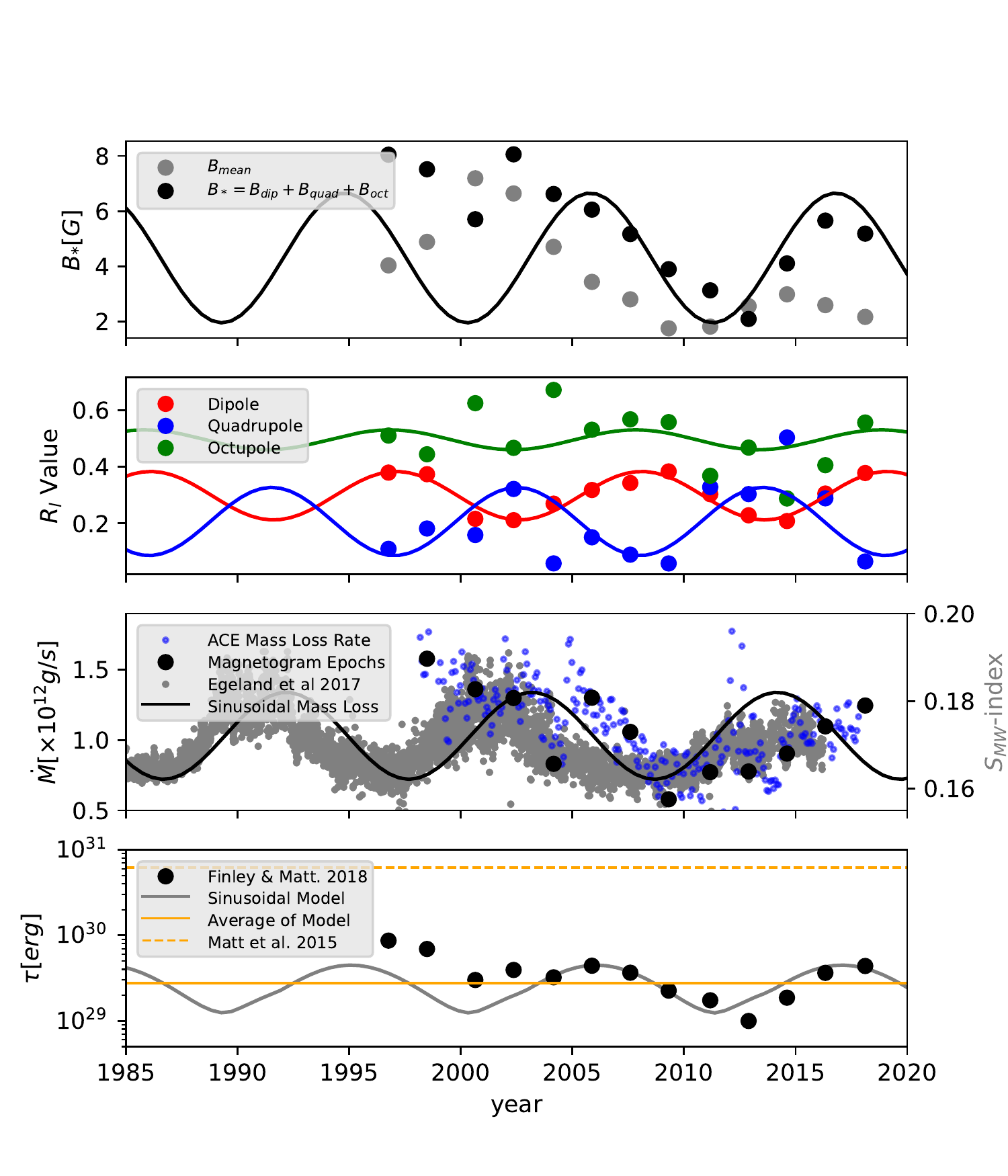}
     \caption{Angular momentum loss calculation for the solar wind (the Sun-as-a-star approach). The top two panels show the magnetic field properties of the Sun using synoptic magnetograms from SOHO/MDI and SDO/HMI. Dots represent sparsely sampled epochs of observation. The first panel shows the evolution of the magnetic field strength at the surface of the Sun. The second panel shows the ratio of dipole, quadrupole and octupole components to the combined (dipole, quadrupole and octupole) magnetic field strength. The third panel displays the mass loss rate measurements derived from the ACE spacecraft (see \citetalias{finley2018effect}) in blue, with the the selected epochs shown with black dots (left scale), along with the evolution of solar S-index from \cite{egeland2017mount} with grey dots (right scale). We fit sinusoids to the magnetic and mass loss rate variables with a fixed 11 year period, which roughly represents the solar chromospheric activity cycle. The fourth panel displays the calculated torques for each magnetogram epoch using \citetalias{finley2018dipquadoct}, with black dots. The torque using our continuous sinusoidal fits is plotted with a solid grey line, and its average is highlighted with a solid orange horizontal line. The torque calculated using \citetalias{matt2015mass} is indicated with a dashed orange horizontal line.}
     \label{sun_figure}
  \end{figure}

\section{Observed Stellar Properties}
We select a sample from all stars that have been monitored with ZDI, requiring that each have 6 or more ZDI observations, which clearly show magnetic variability. This criteria selects 4 stars, as most stars that have been observed with ZDI have only one or two epochs. Along with our sample stars we also consider the Sun. This section contains information on each star, including results from ZDI, studies of their astrospheric Lyman-$\alpha$ absorption, along with proxies of their magnetic activity. Both solar and stellar parameters can be found in Table \ref{StellarParameters}.

\subsection{The Sun}
The study of the Sun's magnetism has afforded the astrophysics community a great wealth of information on the apparent behaviour of the magnetic dynamo process \citep{brun2015recent}. We observe the Sun to have a cyclic pattern in its magnetic activity with a sunspot cycle of around 11 years and a magnetic cycle lasting approximately 22 years \citep{babcock1961topology, schrijver2008global, derosa2012solar}. At the minimum of magnetic activity, the wind dynamics on large scales are dominated by the axisymmetric dipole component and the solar wind is, in general, fast and diffuse, emerging on open polar field lines \citep{wang1990solar, schwenn2006solar}. As the cycle progresses, the solar magnetic field becomes increasingly complex towards maximum, with the appearance of sunspots, as buoyant magnetic flux tubes rise through the photosphere \citep{parker1955formation, spruit1981motion, caligari1995emerging, fan2008three}. Due to the increased complexity, more of the solar wind emerges in the slow, dense component, and transient magnetic phenomena are more frequent \citep{webb1994solar, neugebauer2002sources, mccomas2003three}. The average surface magnetic field is stronger at maximum, and so too are magnetic activity indicators and the solar irradiance \citep{lean1998magnetic, wenzler2006reconstruction}. Following the decline of magnetic activity into the next minimum, the polarity of the field is reversed \citep{babcock1959sun, sun2015polar}. Numerous mechanisms are proposed to explain this \citep[e.g.][]{fisher2000solar, ossendrijver2003solar}. The solar magnetic field returns to its original polarity after one further sunspot cycle, completing the magnetic cycle.

As done in \citetalias{finley2018effect} for the Sun we use synoptic magnetograms taken by the Michelson Doppler Imager on-board the Solar and Heliospheric Observatory (SOHO/MDI) and the Helioseismic and Magnetic Imager on-board the Solar Dynamic Observatory (SDO/HMI). We calculate the average surface magnetic field strength $B_{mean}$, the combined polar dipole, quadrupole and octupole field strength $B_*$, and the field fractions $\mathcal{R}_{dip}$, $\mathcal{R}_{quad}$ and $\mathcal{R}_{oct}$. Unlike \citetalias{finley2018effect}, in order to better compare the solar case with other stars, and to illustrate the effect of sparse time sampling, we take only 13 Carrington rotations, equally spaced over the $\sim 20$ years of data. This information is plotted in the top two panels of Figure \ref{sun_figure} and tabulated in Appendix A.

The first panel of Figure \ref{sun_figure} compares the average surface magnetic field, $B_{mean}$, which is often used when discussing results from ZDI, and the combined polar field strength of the lowest three spherical harmonic components, $B_*$, which is required by the \citetalias{finley2018dipquadoct} torque formulation. $B_*$ is typically larger than $B_{mean}$ as it sums the absolute magnitude of the polar field strengths, whereas $B_{mean}$ allows for opposing field polarities to cancel and is averaged over the stellar surface.

By sparsely sampling the solar magnetograms, the well known cyclic behaviour of the large scale magnetic field has become less obvious, especially when considering $B_*$. However, the cycle is more clear in the second panel, where we plot the fraction of $B_*$ in the dipole, quadrupole and octupole components. We illustratively recover the magnetic behaviour of the Sun by fitting sinusoids of $B_*$, $\mathcal{R}_{dip}$, $\mathcal{R}_{quad}$, and $\mathcal{R}_{oct}$, with a fixed 11 year period, and allowing the phase and amplitude of each fit to vary. These are shown in Figure \ref{sun_figure}, and will be repeated for the ZDI sample in Section 3.3 to produce feasible distributions of magnetic properties for each star, allowing us to further examine the role of magnetic variability on stellar wind torques.

\subsection{Other Stars}
Four stars observed with ZDI meet our criteria for selection, these are {61 Cyg A}, {$\epsilon$ Eri}, {$\xi$ Boo A}, and {$\tau$ Boo A}. Their basic properties are compiled in Table \ref{StellarParameters}. Masses are determined using the stellar evolution model of \cite{takeda2007structure}. If available, radii are either evaluated with interferometry by \cite{kervella2008radii}, \cite{baines2011confirming}, or \cite{boyajian2013stellar}, otherwise spectroscopically by \cite{borsa2015gaps}. Effective temperatures are taken from \cite{boeche2016sp_ace}, which are then used in conjunction with equation (\ref{tcz}) to produce convective turnover timescales. Rotation periods for each star are determined by \cite{saikia2016solar}, \cite{ruedi1997magnetic}, \cite{toner1988starpatch, donahue1996relationship}, and \cite{donati2008magnetic, fares2009magnetic} respectively. These are then used to calculate the Rossby number $R_o=P_{rot}/t_{cz}$ for each object. Further details for each star are listed below:

{61 Cyg A} (HD 201091) is a K5V star, located 3.5 pc away \citep{brown2016gaia} in the constellation of Cygnus as a visual binary with 61 Cyg B, a K7V star. Age estimations for 61 Cyg A range from 1.3 to 6.0 Gyrs, with the majority of estimates on the younger side of this range (2 Gyrs \citealp{barnes2007ages}, 3.6 Gyrs \citealp{mamajek2008improved}, 6 Gyrs \citealp{kervella2008radii}, 1.3 Gyrs \citealp{marsden2014bcool}).  Cyclic chromospheric/coronal activity is detected in many forms including x-ray emission \citep{robrade2012coronal}, with a period in phase with its magnetic activity cycle \citep{baliunas1995chromospheric, saikia2016solar, saikia2018direct}.

{\boldmath$\epsilon$ Eri} (HD 22049) is a K2V star in the constellation of Eridanus, at a distance of 3.2 pc \citep{brown2016gaia}. $\epsilon$ Eri is a young star with multiple age estimations \citep[e.g.][]{song2000ages, fuhrmann2004nearby}. From gyrochronology, \cite{barnes2007ages} arrives at the age of 400Myrs which is thought to be the more reliable \citep[See discussion in][]{janson2008comprehensive}. Chromospheric activity has been recorded for $\epsilon$ Eri by \cite{metcalfe2013magnetic}, displaying an activity cycle length of $\sim3$ years and also a longer of $\sim13$ years which vanished after a 7 year minimum in activity around 1995.

{\boldmath$\xi$ Boo A} (HD 131156A), spectral type G7V, lies in the constellation of Bo$\ddot{ \text{o}}$tes, 6.7 pc away \citep{brown2016gaia} in a visual binary with $\xi$ Boo B of spectral type K5V. The age of $\xi$ Boo A is determined from gyrochronology by \cite{barnes2007ages} as 200Myrs. Variations in $\xi$ Boo A's chromospheric activity are noted by multiple authors \citep{hartmann1979chromospheres, gray1996magnetic, morgenthaler2012long}, but no clear cycle is detected.

{\boldmath$\tau$ Boo A} (HD 120136) is a very well studied planet-hosting F7V star, sitting at a distance of 15.7 pc \citep{brown2016gaia} in a multiple star system with $\tau$ Boo B, a faint M2V companion. $\tau$ Boo A has an age of around 1Gyr \citep{borsa2015gaps}, and has an observed chormospheric activity cycle \citep{mengel2016evolving, mittag2017four}, which is in phase with the reversals of its global magnetic field \cite{jeffers2018relation}. This is also the case for the Sun and 61 Cyg A. As $\tau$ Boo A has a close-in planetary companion. \cite{walker2008most} searched for star-planet interactions and found the planet is likely inducing an active region on the stellar surface causing further variability in the star's chromospheric emission.

\begin{table*}[htbp]
\centering
\caption{Magnetic Properties from ZDI and Angular Momentum Loss Results}
\label{Parameters}
\center
\setlength{\tabcolsep}{3pt}
  \begin{tabular}{c|ccccc|cccc|c}
  \hline\hline
Star	&	ZDI Obs 	&	$B_{*}$	&	$\mathcal{R}_{dip}$	&	$\mathcal{R}_{quad}$	&	$\mathcal{R}_{oct}$	&	$\langle R_A \rangle/R_*$	&	$\tau_{FM18} $	&	$\tau_{spinev} $	&	$\tau_{spinev}$	&	Reference	\\
Name	&	 Epoch	&	 (G)	&	$\equiv B_{dip}/B_{*}$	&	$\equiv B_{quad}/B_{*}$	&	$\equiv B_{oct}/B_{*}$	&		&	 ($\times 10^{30}$erg)	&	 ($\times 10^{30}$erg)	&	$/\langle \tau_{FM18}\rangle$	&	(ZDI data)	\\	\hline
61 Cyg A	&	2007.59	&	17.5	&	0.58	&	0.27	&	0.15	&	11.8	&	0.29	&	5.25	&	26.25	&	1	\\
	&	2008.64	&	4.7	&	0.46	&	0.37	&	0.17	&	5.5	&	0.08	&		&		&	1	\\
	&	2010.55	&	7.3	&	0.28	&	0.29	&	0.43	&	5.1	&	0.09	&		&		&	1	\\
	&	2013.61	&	13.1	&	0.61	&	0.27	&	0.12	&	10.9	&	0.22	&		&		&	1	\\
	&	2014.61	&	11.6	&	0.59	&	0.28	&	0.13	&	9.9	&	0.20	&		&		&	1	\\
	&	2015.54	&	15.6	&	0.65	&	0.25	&	0.10	&	11.4	&	0.32	&		&		&	1	\\	\hline
$\epsilon$ Eri	&	2007.08	&	15.0	&	0.74	&	0.19	&	0.08	&	4.7	&	13.4	&	114	&	11.41	&	2	\\
	&	2007.09	&	15.1	&	0.51	&	0.31	&	0.18	&	4.0	&	9.8	&		&		&	2	\\
	&	2010.04	&	17.1	&	0.36	&	0.37	&	0.27	&	3.5	&	8.2	&		&		&	2	\\
	&	2011.81	&	13.0	&	0.53	&	0.26	&	0.21	&	4.2	&	6.8	&		&		&	2	\\
	&	2012.82	&	20.3	&	0.55	&	0.26	&	0.19	&	4.7	&	14.0	&		&		&	2	\\
	&	2013.75	&	24.6	&	0.66	&	0.16	&	0.18	&	5.6	&	19.1	&		&		&	2	\\
	&	2014.71	&	11.1	&	0.43	&	0.33	&	0.24	&	3.6	&	4.8	&		&		&	3	\\
	&	2014.84	&	11.6	&	0.53	&	0.23	&	0.24	&	4.0	&	6.2	&		&		&	3	\\
	&	2014.98	&	13.7	&	0.54	&	0.27	&	0.19	&	4.2	&	7.6	&		&		&	3	\\	\hline
$\xi$ Boo A	&	2007.56	&	42.8	&	0.56	&	0.24	&	0.20	&	11.0	&	29.1	&	748	&	32.4	&	4	\\
	&	2008.09	&	32.3	&	0.45	&	0.27	&	0.27	&	8.5	&	20.0	&		&		&	4	\\
	&	2009.46	&	42.4	&	0.42	&	0.29	&	0.29	&	8.8	&	27.3	&		&		&	4	\\
	&	2010.04	&	24.1	&	0.48	&	0.27	&	0.25	&	7.9	&	14.8	&		&		&	4	\\
	&	2010.48	&	37.8	&	0.53	&	0.27	&	0.20	&	9.8	&	26.6	&		&		&	4	\\
	&	2010.59	&	24.5	&	0.47	&	0.29	&	0.24	&	7.4	&	16.8	&		&		&	4	\\
	&	2011.07	&	26.5	&	0.60	&	0.26	&	0.14	&	8.1	&	27.0	&		&		&	4	\\	\hline
$\tau$ Boo A	&	2008.04	&	2.2	&	0.33	&	0.33	&	0.35	&	2.1	&	108	&	367	&	2.72	&	5	\\
	&	2008.54	&	1.8	&	0.33	&	0.33	&	0.34	&	2.0	&	141	&		&		&	5	\\
	&	2008.62	&	1.8	&	0.32	&	0.36	&	0.32	&	2.0	&	133	&		&		&	5	\\
	&	2009.5	&	2.5	&	0.39	&	0.33	&	0.28	&	2.1	&	156	&		&		&	5	\\
	&	2010.04	&	3.0	&	0.35	&	0.35	&	0.30	&	2.2	&	109	&		&		&	6	\\
	&	2011.04	&	2.7	&	0.48	&	0.23	&	0.28	&	2.1	&	127	&		&		&	6	\\
	&	2011.45	&	2.5	&	0.22	&	0.38	&	0.40	&	2.1	&	163	&		&		&	6	\\
	&	2013.45	&	3.1	&	0.34	&	0.34	&	0.32	&	2.2	&	142	&		&		&	6	\\
	&	2013.96	&	3.8	&	0.41	&	0.39	&	0.20	&	2.2	&	170	&		&		&	6	\\
	&	2014.45	&	2.5	&	0.34	&	0.31	&	0.35	&	2.1	&	108	&		&		&	6	\\
	&	2015.04	&	2.9	&	0.35	&	0.31	&	0.34	&	2.2	&	146	&		&		&	6	\\
	&	2015.29	&	1.6	&	0.59	&	0.24	&	0.17	&	1.9	&	141	&		&		&	6	\\
	&	2015.33	&	1.3	&	0.58	&	0.26	&	0.16	&	1.8	&	123	&		&		&	6	\\
	&	2015.35	&	1.6	&	0.58	&	0.24	&	0.18	&	1.9	&	123	&		&		&	6	\\
	&	2015.38	&	2.4	&	0.45	&	0.28	&	0.27	&	2.1	&	124	&		&		&	6	\\
	&	2016.21	&	3.2	&	0.49	&	0.27	&	0.24	&	2.2	&	166	&		&		&	7	\\
	&	2016.44	&	2.1	&	0.29	&	0.33	&	0.38	&	2.1	&	97	&		&		&	7	\\
	&	2016.47	&	3.0	&	0.44	&	0.25	&	0.31	&	2.2	&	124	&		&		&	7	\\
	&	2016.54	&	2.7	&	0.42	&	0.29	&	0.29	&	2.1	&	160	&		&		&	7	\\
	\hline
  \end{tabular}
  \item{\small [1] \cite{saikia2016solar}, [2] \cite{jeffers2014e}, [3] \cite{jeffers2017relation}, [4] \cite{morgenthaler2012long},}
  \item{\small [5] \cite{fares2009magnetic}, [6] \cite{mengel2016evolving}, [7] \cite{jeffers2018relation}. }
\end{table*}

\subsection{Zeeman-Doppler Imaged Fields}
{61 Cyg A} \citep{saikia2016solar}, {$\epsilon$ Eri} \citep{jeffers2014e, jeffers2017relation}, {$\xi$ Boo A} \citep{morgenthaler2012long}, and {$\tau$ Boo A} \citep{fares2009magnetic, mengel2016evolving, jeffers2018relation} have all been monitored with ZDI. This is a tomographic technique that is capable of reconstructing their large-scale photospheric magnetic fields \citep{semel1989zeeman, donati1989zeeman, brown1991zeeman, donati1997zeeman, donati2009magnetic}. Magnetic fields cause spectral lines to split and become polarized due to the Zeeman effect \citep{zeeman1897effect}. By monitoring this splitting over multiple phases, taking advantage of the doppler shifts due to rotation, and combining multiple line profiles together using a Least Squares Deconvolution (LSD) technique \citep{donati1997spectropolarimetric}, the large-scale stellar magnetic field topology can be reconstructed.

Papers reporting ZDI results typically tabulate the farction of the total magnetic field energy that is poloidal ($E_{pol}$) and the farction of this poloidal field energy that is dipolar, quadrupolar or octupolar ($E_{dip}$, $E_{quad}$, and $E_{oct}$), and the average surface field ($B_{mean}$). For the maps of \cite{fares2009magnetic} and \cite{mengel2016evolving} we compute the values using data supplied by the authors, since these values are not tabulated in the original papers. Using MHD stellar wind models, \cite{jardine2013influence} were able to show that large scale wind dynamics are largely unaffected by toroidal magnetic field structures embedded in the photosphere. Therefore we assume the toroidal component does not impact our torque calculations. We convert the percentage energies, into the poloidal dipole, quadrupole and octupole field fractions, and combined field strength,
\begin{eqnarray}
f_{dip}&=&\sqrt{E_{pol}E_{dip} },\\
f_{quad}&=&\sqrt{E_{pol}E_{quad}  },\\
f_{oct}&=&\sqrt{E_{pol}E_{oct}  },\\
B_*&=&B_{mean}(f_{dip}+f_{quad}+f_{oct}).
\end{eqnarray}
Here care has been taken in transforming fractional energy into fractional field strengths for each magnetic component. Subsequently, the field fractions, $f_{dip}, f_{quad}$, and $f_{oct}$ are converted into the ratios of each magnetic component to the combined field strength, $\mathcal{R}_l$,
\begin{eqnarray}
\mathcal{R}_{dip}&=&\frac{{f_{dip}}}{{f_{dip}}+{f_{quad}}+{f_{oct}}} \equiv \frac{B_{dip}}{B_*},\\
\mathcal{R}_{quad}&=&\frac{{f_{quad}}}{{f_{dip}}+{f_{quad}}+{f_{oct}}} \equiv \frac{B_{quad}}{B_*},\\
\mathcal{R}_{oct}&=&\frac{{f_{oct}}}{{f_{dip}}+{f_{quad}}+{f_{oct}}} \equiv \frac{B_{oct}}{B_*}.
\end{eqnarray}
These results are shown in the top two panels of each Figure \ref{61Cyg_figure}-\ref{tauBoo_figure}, and tabulated in Table \ref{Parameters} for each ZDI epoch. {Calculating the ratios of each field component using this method, rather than re-computing the field strengths of each component from the original ZDI maps, introduces some error which will be discussed in Section 5.1.}

The first panel for each Figure \ref{61Cyg_figure}-\ref{tauBoo_figure} displays the recorded mean magnetic field from the ZDI reconstructions, $B_{mean}$ with grey dots. The black dots represent the combined polar field strength of the dipole, quadrupole and octupole components, $B_*$. Typically the $B_*$ value is larger than $B_{mean}$, unless a significant fraction of the magnetic energy is stored in the toroidal or high order ($l>3$) components. The second panels show the varying field fractions, $\mathcal{R}_{dip}$, $\mathcal{R}_{quad}$, and $\mathcal{R}_{oct}$.

Although multiple magnetic maps exist for each of our ZDI stars, they are still relatively sparsely sampled compared to the Sun. To examine their variability further, we fit sinusoidal functions to $B_*$, $\mathcal{R}_{dip}$, $\mathcal{R}_{quad}$, and $\mathcal{R}_{oct}$, as we did for the Sun, using chromospheric activity periods taken from the literature for each star (see Table \ref{StellarParameters}). We allow the phase and amplitude of each fit to vary, however we constrain the fits of $\mathcal{R}_{dip}$, $\mathcal{R}_{quad}$, and $\mathcal{R}_{oct}$ to sum to $\sim 1$. In some cases there is no strong evidence for periodicity and even if so, a sinusoidal behaviour is a gross simplification. We do this simply to illustratively construct continuous predictions for feasible cyclic behaviours, from which, we can make more general comments about the impact of stellar cycles on stellar wind torques.

\subsection{Inferred Mass Loss Rates and Activity Proxies}
The solar mass loss rate is observed to be variable in time \citep{hick1994solar, webb1994solar, mccomas2000solar, mccomas2013weakest}. In the third panel of Figure \ref{sun_figure}, we plot the solar mass loss rate calculated in \citetalias{finley2018effect}, using data from the Advanced Composition Explorer\footnote{http://srl.caltech.edu/ACE/ASC/level2/} with blue dots, and highlight our selected magnetogram epochs with black dots. {During the solar cycle, the mass loss rate from \citetalias{finley2018effect} is found to vary around the mean by around $\pm30\%$.\footnote{In calculating this variation, we ignore extreme values that are seen in time averages shorter than a few months.}. We fit the function,
\begin{equation}
  \dot M(t) = \langle \dot M \rangle \bigg[0.3\sin\bigg(\frac{2\pi t}{P}+\phi\bigg) + 1\bigg],
\end{equation}
to the 13 selected magnetogram epochs. Where $t$ is the decimal year (1985 to 2020 is plotted), the mass loss rate variation is constrained to $\Delta \dot M = 0.6\langle \dot M\rangle$, and the period is fixed that of the chromospheric activity period, $P = 11$ years. The fit values of the phase, $\phi$ and the average mass loss rate, $\langle \dot M \rangle$, are $\sim\pi/6$ and $1.03\times 10^{12}$g/s respectively.} This fit is shown in the third panel with a solid black line.

For nearby stars, Lyman-$\alpha$ observations can reveal information about their stellar winds \citep{wood2004astrospheres}. Absorption in this line occurs at the edge of the star's astropshere as well as at the Sun's heliosphere. At these locations, the solar and stellar winds collide with the ISM and become shocked, reaching temperatures and densities much greater than the average ISM. Through modelling this absorption, estimated mass loss rates are available from \cite{wood1998local}, \cite{wood2002measured}, and \cite{wood2005new}, for 61 Cyg A, $\epsilon$ Eri, and $\xi$ Boo A. For $\tau$ Boo A, there are no measurements of the mass loss rate so instead we use the results of MHD simulations from \cite{nicholson2016temporal}. The mass loss rate used for each star is shown in Table \ref{StellarParameters}

For the ZDI stars, the mass loss rates gathered from Lyman-$\alpha$ observations are taken at a single epoch. These are plotted as black dots in the third panel of Figures \ref{61Cyg_figure}-\ref{xiBoo_figure}. However, we might expect the mass loss rates of these stars to vary with their magnetic activity similarly to the Sun. Currently their are no observations in the literature capable of quantifying this variability, therefore we must draw comparisons with the Sun.

Increased emission in Ca II H\&K is thought to correlate directly with the deposition of magnetic energy into the stellar chromosphere \citep{eberhard1913reversal, noyes1984rotation, testa2015stellar}. This is observed for the Sun \citep{schrijver1989relations} and can be correlated with the solar wind mass loss rate. Over-plotted with the mass loss rates in Figure \ref{sun_figure}, we show the solar S-index values from \cite{egeland2017mount}. The S-index evaluates the flux in the H and K lines and normalises it to the nearby continuum \citep{wilson1978chromospheric}. The solar mass loss rate, and the sinusoidal fit to our selected epochs, both appear roughly in phase with this measure of chromospheric activity. The slight lag between mass loss rate and magnetic activity is not surprising, as a similar lag is observed in the rate of coronal mass ejections \citep{ramesh2010coronal, webb2012coronal}, and open magnetic flux in the solar wind \citep{wang2000long,owens2011open}. The Ca II H\&K lines are now regularly monitored for hundreds of stars \citep{wilson1978chromospheric, baliunas1995chromospheric, hall2007activity, egeland2017mount}. We plot the available S-index measurements for each star in the third panel of Figures \ref{61Cyg_figure}-\ref{tauBoo_figure} with grey dots. The temporal coverage differs from star to star, with $\xi$ Boo A having only the Ca II H band index\footnote{As both the H and K lines scale together, only information about one is required.}, taken concurrently with the ZDI observations \citep{morgenthaler2012long}.

{Similarly to the Sun, we represent the mass loss variation for each star using a sinusoidal function,
\begin{equation}
  \dot M(t) = 0.3 \langle \dot M \rangle \bigg[\sin\bigg(\frac{2\pi t}{P}+\phi\bigg)-\sin\bigg(\frac{2\pi t_{obs}}{P}+\phi\bigg)\bigg] + \dot M_{obs},
\end{equation}
with the phase, $\phi$, and period, $P$, matching the variation of their Ca II H\&K emission. We use chromospheric activity periods from the existing literature (see Table \ref{StellarParameters}), and show the available Ca II H\&K indices in Figures \ref{61Cyg_figure}-\ref{xiBoo_figure}. Although a correlation between mass loss rate and Ca II H\&K emission seems to exist for the Sun (visible in Figure \ref{sun_figure}), the correlation is complex, and it is not obvious whether a similar relationship exists for other stars. {If we were to use the correlation for the Sun to estimate the mass loss rate variation of our sample stars, given their variability in Ca II H\&K emission, i.e. $\Delta \dot M \propto \Delta S_{index}$, we would find a range of amplitudes around $\Delta \dot M = 0.01-1.5\langle \dot M\rangle$.} Given the uncertainties, we simply adopt the same amplitude for the mass loss rate as was determined for the Sun ($\Delta \dot M = 0.6\langle \dot M\rangle$), and require the function to reproduce the astropheric Lyman-$\alpha$ observations (i.e., $\dot M(t_{obs})=\dot M_{obs}$). The solid black line in each Figure represents this projected variability. Note that since the torque is a relatively weak function of mass loss rate (see equations (\ref{torque})-(\ref{DQO_law})), our assumption about the amplitude of variability in mass loss rate has a similarly weak effect on the amplitude of variability in the torque.}

\section{Angular Momentum Loss Rates}
Here we apply the \citetalias{finley2018dipquadoct} braking law to our sample stars to calculate their stellar wind torques. We also calculate the rotational evolution torques from \citetalias{matt2015mass}.

\subsection{Predicted Alfv\'en Radii}
Through the application of \citetalias{finley2018dipquadoct} to our sample stars, we are able to examine their individual locations in our MHD parameter space. The location of each ZDI epoch and sinusoidal model in $\langle R_A\rangle - \Upsilon$ space, are displayed in Figure \ref{RA_figure}. {Uncertainties in the recovered field strengths from ZDI are difficult to quantify. Typically, errors quoted in ZDI papers are obtained by varying the input parameters to reconstruct additional ZDI maps, from which the variation in field strengths are quoted as error \citep[discussed in][]{petit2008toroidal}. We propagate typical uncertainties for the each magnetic field strength ($\pm30\%$), and the mass loss rates ($\pm50\%$), using standard error analysis.} The resulting uncertainty in wind magnetisation, $\Upsilon$, and the average Alfv\'en radius, $\langle R_A\rangle$, are correlated which we show with diagonal grey lines in Figure \ref{RA_figure}. Vertical lines represent a $\pm10\%$ uncertainty on our prediction of $\langle R_A\rangle$, which considers the approximations made in fitting equation (\ref{DQO_law}). This is discussed further in \citetalias{finley2018dipquadoct} (see their Figure 10).

The wind magnetisation parametrises the effectiveness of the wind braking, or more physically, the size of the torque-averaged Alfv\'en radius. However equation (\ref{DQO_law}) also encodes information about the magnetic geometry of the field, approximating this effect as a twice broken power law. Depending on the strength of the three magnetic geometries considered here, the dipolar, quadrupolar, or octupolar (top, middle or bottom) formula in equation (\ref{DQO_law}) will be used to calculate $\langle R_A\rangle$. To identify when each formula is used, different symbols are plotted in Figure \ref{RA_figure}.

The average Alfv\'en radii of our sample stars range from $\sim2-11R_*$, most being typically dipole dominated with the exception of $\tau$ Boo A. The predicted $\langle R_A\rangle$ values for $\tau$ Boo A follow a shallower slope than the other dipolar dominated stars, due to the weaker dependence of the octupolar geometry, compared with the dipole or quadrupole geometries, on wind magnetisation in equation (\ref{DQO_law}). The MHD model results of \cite{nicholson2016temporal} for the $\langle R_A\rangle$ of $\tau$ Boo A, are also plotted with light blue squares in Figure \ref{RA_figure}. Their values for $\langle R_A\rangle$ are shown to be in good agreement with results from the \citetalias{finley2018dipquadoct} braking law.

The Sun appears typical when compared with the three dipole dominated stars, with some having larger $\langle R_A\rangle$ and some having smaller. However, the Sun shows some quadrupolar dominated behaviour around solar maximum, which is not observed in the other dipole dominated stars. Each sinusoidal model roughly represents the observed epochs from ZDI, and is able to show how sub-sampling may skew our perception of where each star lies in this parameter space. A similar representation of the solar cycle in this parameter space was explored in the work of \cite{pinto2011coupling} (see Figure 11 within). We find (though not shown) the sinusoidal prediction for the location of the Sun in this parameter space is representative of using the full dataset examined in \citetalias{finley2018effect}.

\subsection{Torques}
\subsubsection{The Sun-as-a-Star}

In \citetalias{finley2018effect} we produced an estimate for the solar angular momentum loss rate using the wealth of observations available for our closest star. Here we instead treat the Sun as a star by reducing the number of observations to approximately $2$ year intervals, thus illustrating the effect of sparse time-sampling. Details on the selected magnetogram epochs are tabulated in Appendix A.

Figure \ref{sun_figure} shows the result of our angular momentum loss calculation. For the Sun, the dipole and octupole geometries are shown to cycle in phase, with the quadrupole out of phase, as previously discussed in \cite{derosa2012solar}. The S-index values from \cite{egeland2017mount} appear in phase with the quadrupolar geometry, and the mass loss rates taken from \citetalias{finley2018effect}. The torques for each epoch using \citetalias{finley2018dipquadoct} are plotted in the bottom panel with black dots. A grey line indicates the torque using the sinusoidal fits of the magnetic field and mass loss rate.

From Figure \ref{sun_figure}, it is clear that simple sinusoids with fixed amplitude and phase are a poor fit to the data. This is primarily due to cycle to cycle variation, i.e. the length of the Sun's magnetic cycle is know to vary, along with the strength of each cycle \citep[e.g.][]{solanki2002search}. However, the poor fit is also representative of the effects of sparse sampling on a system which contains variability on much shorter timescales than considered. Therefore, when considering the magnetic behaviour of other stars, we expect not to see clear cyclical behaviours, even if the stars are truly cyclical, like we know the Sun to be.

We calculate the average torque for the solar magnetogram epochs to be $0.37\times10^{30}$ erg, which is in close agreement with the estimate produced in \citetalias{finley2018effect}. The sinusoidal fits produce an average torque of $0.30\times10^{30}$ erg. The model torque has a different phase with respect to the solar magnetic cycle, than using the full dataset in \citetalias{finley2018effect}, which is a consequence of fitting to sparsely sampled data. The torque given by \citetalias{matt2015mass} is $6.2\times10^{30}$ erg. The discrepancy between these torques is discussed in Section 4.3.

\subsubsection{61 Cygni A}
\begin{figure}
   \centering
    \includegraphics[trim=0cm 0cm 0cm 0cm, width=.49\textwidth]{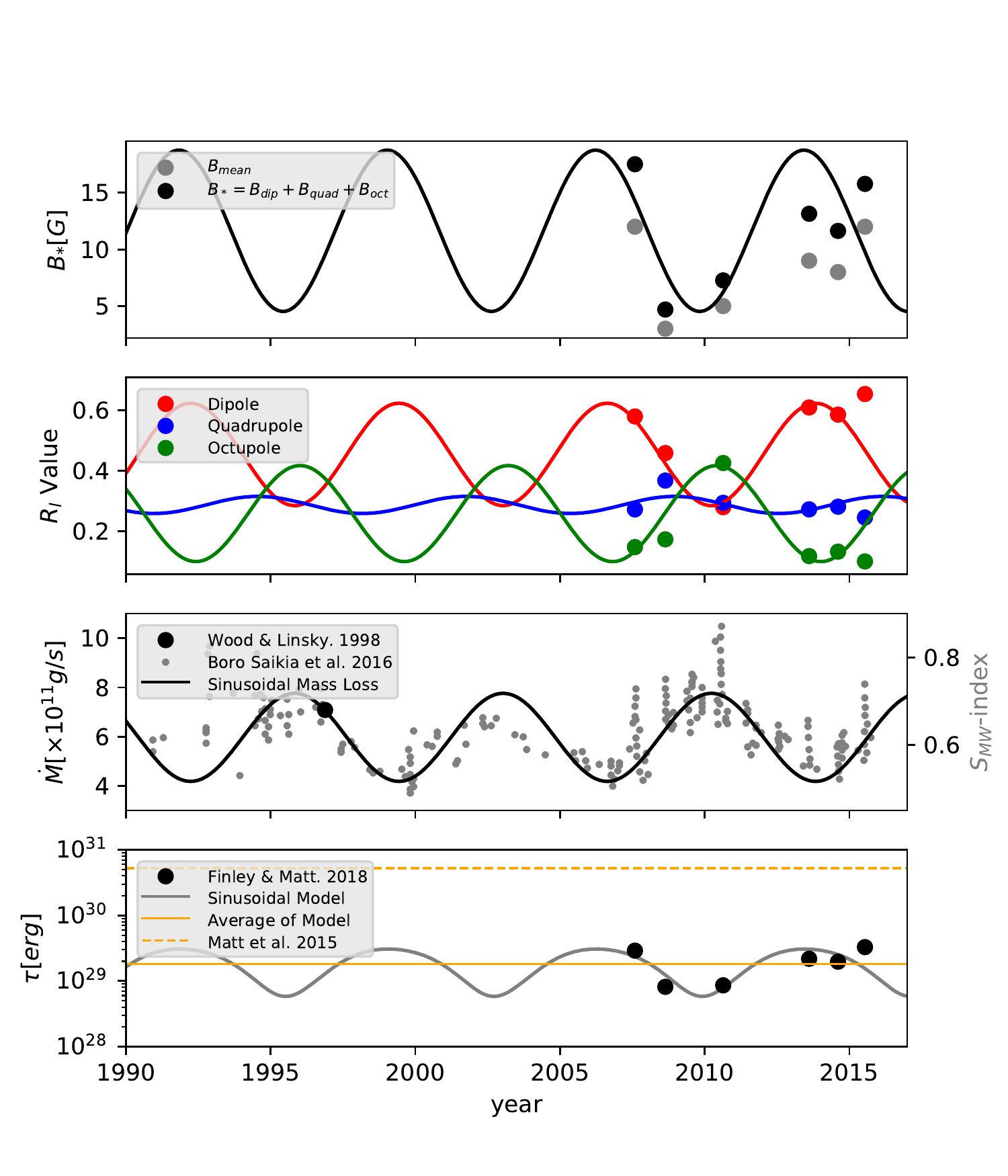}
     \caption{Angular momentum loss calculation for 61 Cyg A. The top two panels show the magnetic field properties taken from the ZDI measurements of \cite{saikia2016solar}. The first panel shows the evolution of the average unsigned magnetic field strength, and the combined (dipole, quadrupole and octupole) magnetic field strength, at the surface of the star. The second panel shows the ratios of dipole, quadrupole and octupole components of the magnetic field to the combined magnetic field strength. We fit sinusoids to these properties with a fixed period of 7.3 years, matching the chromospheric activity cycle. The third panel displays the mass loss rate measurement of \cite{wood1998local} with a black dot, along with the S-index evolution of the chromospheric activity with grey dots \citep{saikia2016solar}. A sinusoidal mass loss rate with a solar-like amplitude, and with phase and period matching the observed chromospheric activity is shown with a solid black line. The fourth panel displays the calculated torques for each ZDI epoch using \citetalias{finley2018dipquadoct} with black dots. The torque using our continuous sinusoidal fits are plotted with a solid grey line, with its average highlighted by a solid orange horizontal line. The torque calculated using \citetalias{matt2015mass} is indicated with a dashed orange horizontal line. }
     \label{61Cyg_figure}
\end{figure}

61 Cyg A was observed with ZDI by \cite{saikia2016solar} from 2007.59 to 2015.54 with an average of 1.19 years between observations. They find a star very much like the Sun in its magnetic behaviour, having both the poloidal and toroidal field components reverse polarity in phase with its chromospheric activity, and a weak solar-like differential rotation profile. Like the Sun, the global field is strongly dipolar with the dipole component strengthening at activity minimum and weakening at activity maximum in favour of more multipolar field geometries.

Figures \ref{61Cyg_figure} and \ref{RA_figure} display the full results of our angular momentum loss calculation. In the bottom panel of Figure \ref{61Cyg_figure}, the values of the torque calculated for the individual ZDI epochs using the projected mass loss rates, are plotted with black dots. The sinusoidal model torque is plotted with a solid grey line. With activity minima in 2007 and 2014, the dipole component is strong and so we predict a large average Alfv\'en radius ($\sim10R_*$, see Figure \ref{RA_figure}). At the activity maximum around 2010, the field is at its most complex. However, the magnetic braking is still dominated by the dipolar component due to the relative strengths of the other modes. This produces the smallest average Alfv\'en radius ($\sim5R_*$).

The average torque for the ZDI epochs of 61 Cyg A, using \citetalias{finley2018dipquadoct}, is $0.20\times 10^{30}$ erg. The average of the sinusoidal model has a similar value of $0.18\times 10^{30}$ erg. The torque from \citetalias{matt2015mass} is calculated to be $5.25\times 10^{30}$ erg.

\subsubsection{$\epsilon$ Eridani}
\begin{figure}
   \centering
    \includegraphics[trim=0cm 0cm 0cm 0cm, width=.49\textwidth]{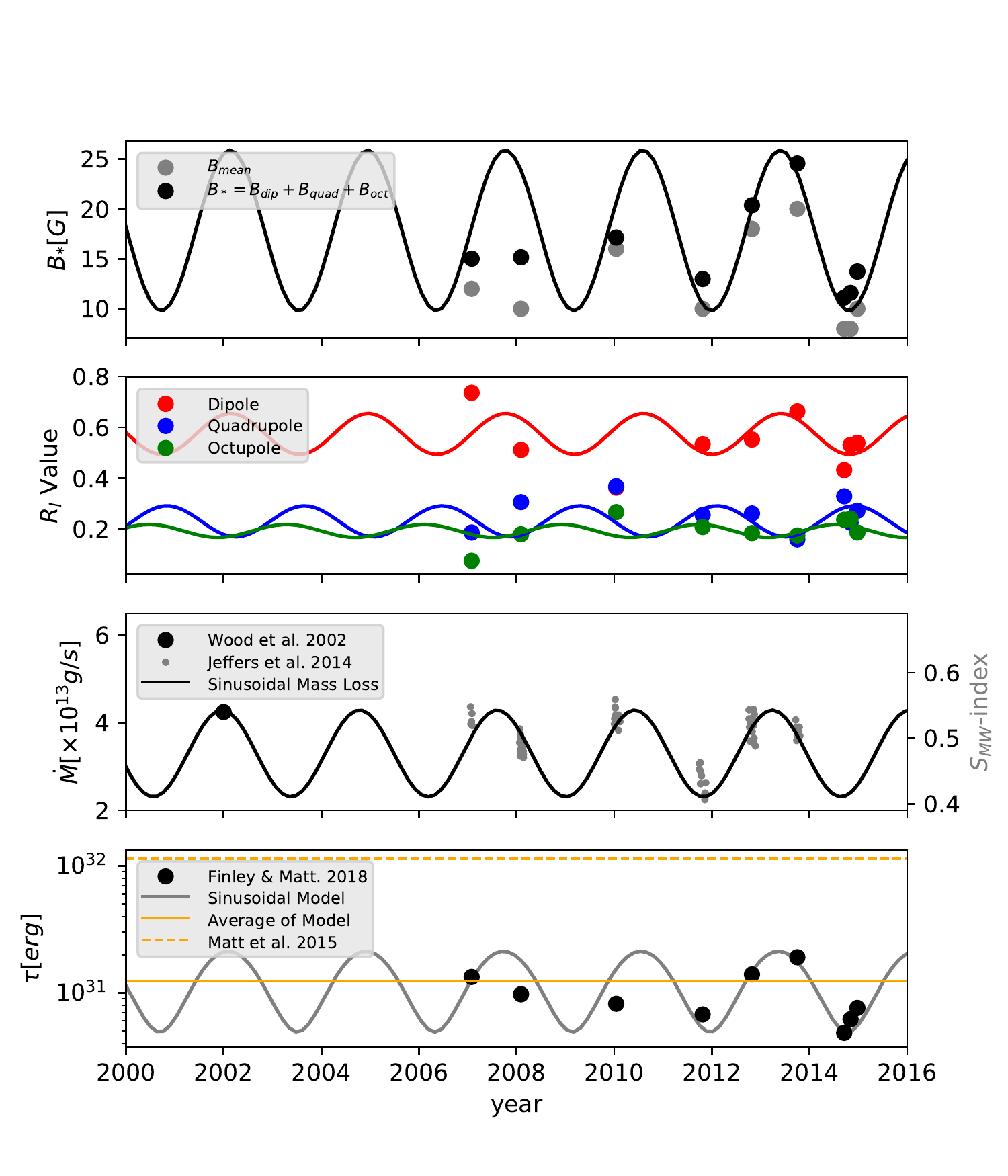}
     \caption{Same as Figure \ref{61Cyg_figure}, now for $\epsilon$ Eri.}
     \label{epEri_figure}
\end{figure}

$\epsilon$ Eri was observed with ZDI by \cite{jeffers2014e} from 2007.08 to 2014.98. \cite{jeffers2014e} originally monitored $\epsilon$ Eri with an average of 1.11 years between observations until 2013.75. \cite{jeffers2017relation} followed up these observations taking 3 observations in quick succession (approximately once a month) during its activity minimum. The magnetic geometry of $\epsilon$ Eri at minimum activity is more complicated than the axisymmetric dipolar structure seen from the Sun and 61 Cyg A. The dipole component instead strengthens at activity maxima, producing the largest Alfv\'en radii when the chromospheric activity is highest.  Figure \ref{epEri_figure} details the angular momentum loss calculation for $\epsilon$ Eri, and the average Alfv\'en radii are displayed in Figure \ref{RA_figure}.

The average torque for the ZDI epochs of $\epsilon$ Eri, using \citetalias{finley2018dipquadoct}, is $1.00\times 10^{31}$ erg. With the sinusoidal fits we find a larger average value of $1.24\times 10^{31}$ erg. The sinusoidal model suggests that the ZDI epochs have preferentially sampled minima of activity, and therefore average to a lower torque. We calculate the torque using \citetalias{matt2015mass} and find a value of $1.14\times 10^{32}$ erg.

\subsubsection{$\xi$ Bootis A}
\begin{figure}
   \centering
    \includegraphics[trim=0cm 0cm 0cm 0cm, width=.49\textwidth]{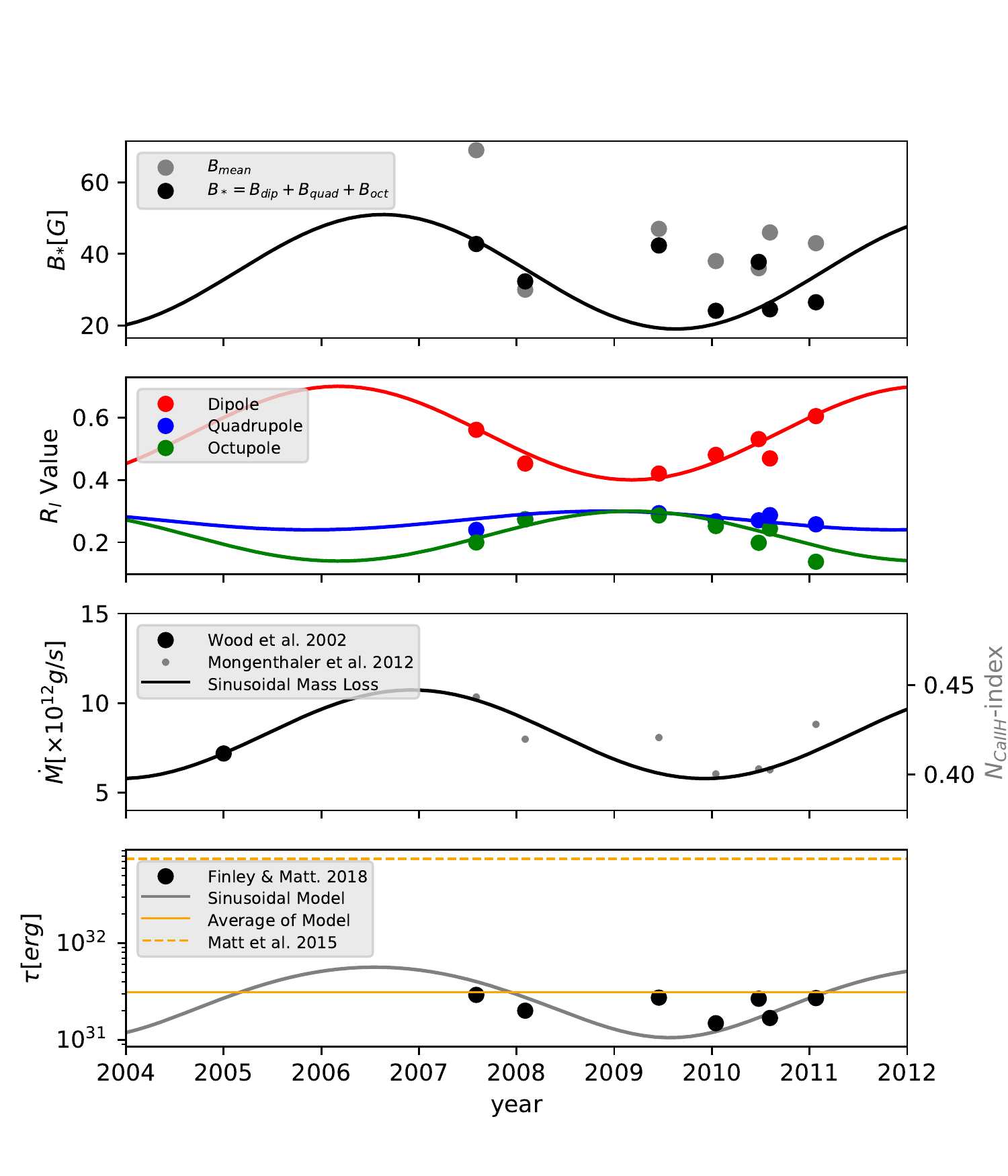}
     \caption{Same as Figure \ref{61Cyg_figure}, now for $\xi$ Boo A.}
     \label{xiBoo_figure}
\end{figure}

The magnetic variability of $\xi$ Boo A is unlike both 61 Cyg A and $\epsilon$ Eri. It was observed with ZDI by \cite{morgenthaler2012long} from 2007.59 to 2011.07, with an average time between observations of half a year. The star hosts a persistent toroidal component with fixed polarity through all observations. This field contains a large fraction of the magnetic energy, shown by the mean field strength (grey dots) in the top panel of Figure \ref{xiBoo_figure} being much larger than the combined magnetic field strength (black dots). The total magnetic field appears to have short time variability. However, the second panel in Figure \ref{xiBoo_figure} appears to show a coherent pattern. With the limited data available, and no cyclic variability detected in other activity indicators, we fit a sinusoid to this slowly varying magnetic geometry.

Note that the data is best represented with maxima occurring where there are no data. The existence and amplitude of the fit maxima is poorly constrained by the available data, and the sinusoidal fit is merely speculative. This leads to the torque for the cycle, shown with a solid grey line in the bottom panel of Figure \ref{xiBoo_figure}, to be much larger than the ZDI epochs, shown with black dots.

The average torque calculated for the ZDI epochs of $\xi$ Boo A, using \citetalias{finley2018dipquadoct}, is $2.31\times 10^{31}$ erg. Averaging the sinusoidal model instead, we produce a torque of $3.10\times 10^{31}$ erg. The rotational evolution torque from \citetalias{matt2015mass} gives a value of $7.48\times 10^{32}$ erg.
\subsubsection{$\tau$ Bootis A}
\begin{figure}
   \centering
    \includegraphics[trim=0cm 0cm 0cm 0cm, width=.49\textwidth]{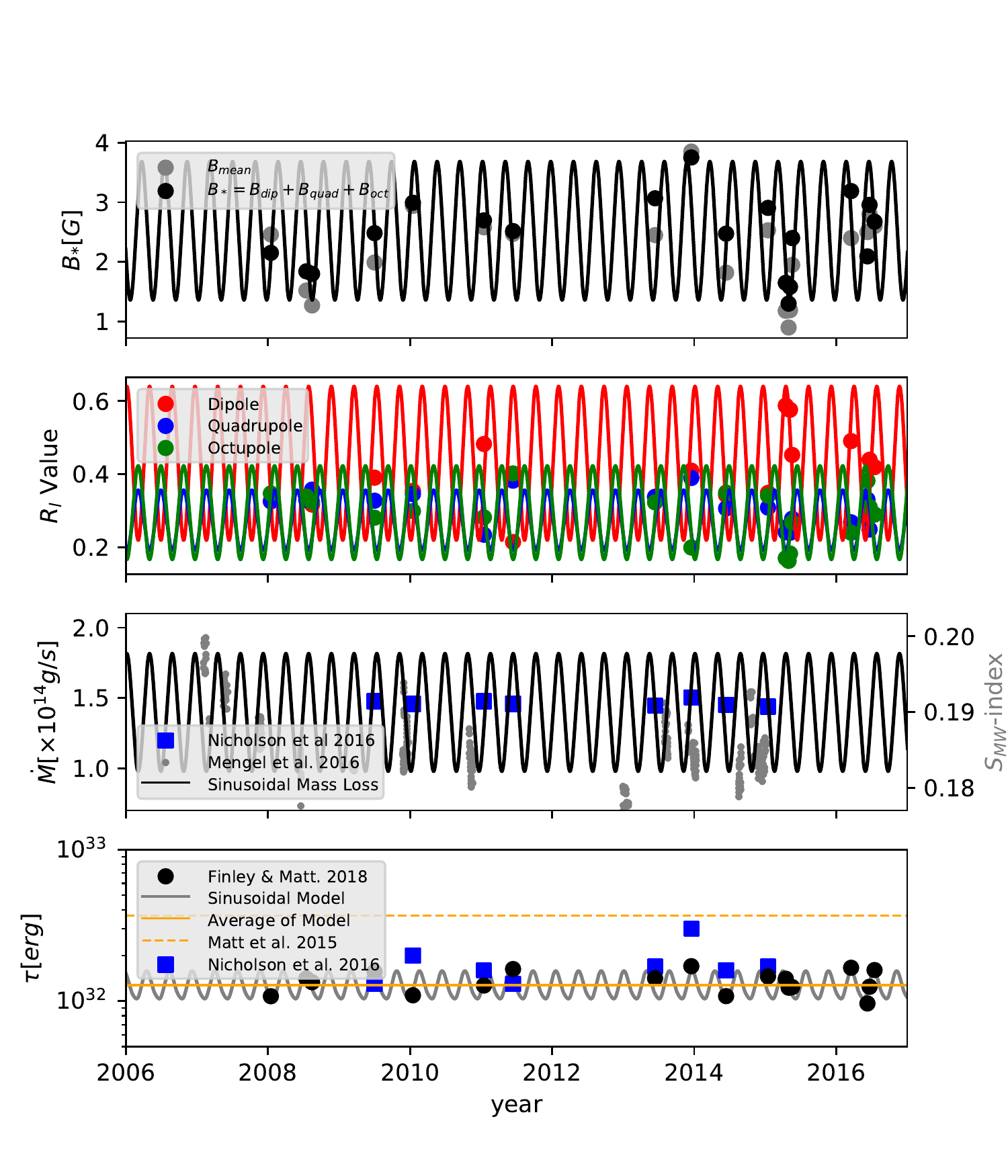}
     \caption{Same as Figure \ref{61Cyg_figure}, now for $\tau$ Boo A. Mass loss rate and torque panels (3 and 4) include values (blue squares) from the MHD simulations of $\tau$ Boo A from \cite{nicholson2016temporal}. A phase-folded version of this plot is available in Appendix B.}
     \label{tauBoo_figure}
\end{figure}

\begin{figure*}
   \centering
    \includegraphics[trim=2cm 0cm 2cm 0cm, width=\textwidth]{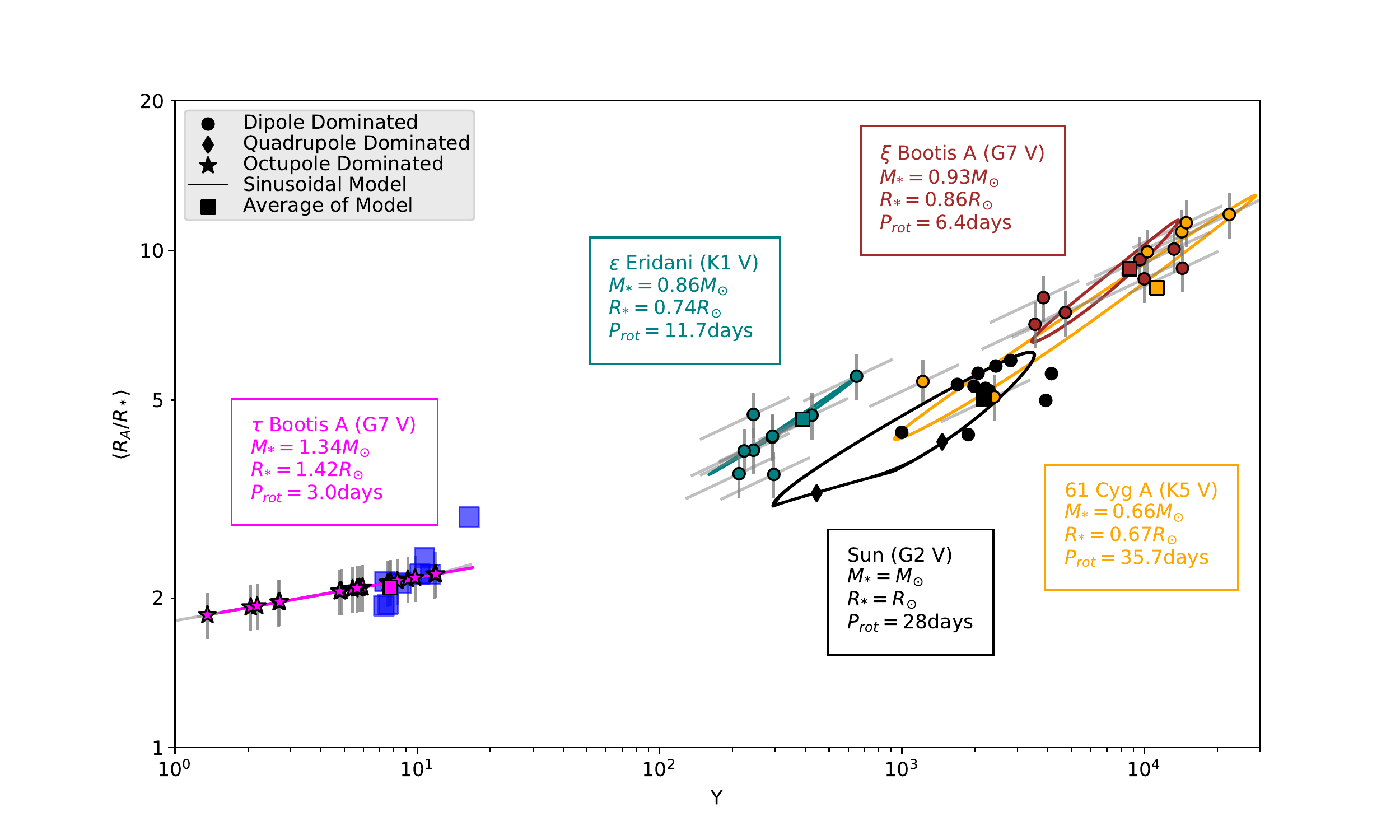}
     \caption{Average Alfv\'en radius versus wind magnetisation, $\Upsilon$. Results for the Sun are shown in black. The ZDI epochs for 61 Cyg A (orange), $\epsilon$ Eri (teal), $\xi$ Boo A (brown) and $\tau$ Boo A (magenta) are displayed with their uncertainties, in grey. The shape of each point signifies the magnetic geometry governing the angular momentum loss rate according to equation (\ref{DQO_law}). Dipolar dominated with circles, quadrupolar dominated with diamonds and octupolar dominated with stars. The sinusoidal models are shown with a corresponding colored line. The average of both quantities for each star are marked with colored squares. The majority of ZDI epochs and solar magnetograms are dominated by the dipolar component, with the exception of $\tau$ Boo A which host a weakly magnetised wind (according to the predictions of $\dot M$ of \cite{nicholson2016temporal}) and so is dominated by the octupolar term in equation (\ref{DQO_law}). Results from the 3D MHD simulations of $\tau$ Boo A from \cite{nicholson2016temporal} are displayed using blue squares, in good agreement with this work.}
     \label{RA_figure}
\end{figure*}

$\tau$ Boo A is currently the most extensively monitored star with ZDI \citep{donati2008magnetic, fares2009magnetic, mengel2016evolving, jeffers2018relation}. From these studies, authors have found $\tau$ Boo A to have a magnetic cycle with polarity reversals in phase with its chromospheric activity cycle of 120 day, as observed for the Sun and 61 Cyg A. Its mass loss rate is not observationally constrained, but MHD simulations of the stellar wind surrounding $\tau$ Boo A have been produced by \cite{nicholson2016temporal}, using maps from some of the ZDI epochs considered here. We include these results in Figure \ref{tauBoo_figure} using blue squares to indicate their derived mass loss rates and angular momentum loss rates. We calculate the torque-average Alfv\'en radii associated with these simulated values using equation (\ref{torque}), and include them in Figure \ref{RA_figure} with light blue squares. For clarity, we also show a phase-folded version of Figure \ref{tauBoo_figure} in Appendix B.

Equation (\ref{DQO_law}) predicts the efficiency of angular momentum loss to be low, and dominated by the octupolar scaling. Both this work and the simulations of \cite{nicholson2016temporal} predict a torque-averaged lever arm of $\sim2R_*$, which is much lower than the other stars in the sample (see Figure \ref{RA_figure}). We calculate an average torque from the ZDI epochs of $\tau$ Boo A, using \citetalias{finley2018dipquadoct}, as $1.23\times 10^{32}$ erg. The sinusoidal model has an average torque of $1.32\times 10^{32}$ erg. The torque from \citetalias{matt2015mass} is calculated to be $3.67\times 10^{32}$ erg.

\begin{figure}
  \centering
   \includegraphics[trim=1cm 0cm 1cm 0cm,width=0.48\textwidth]{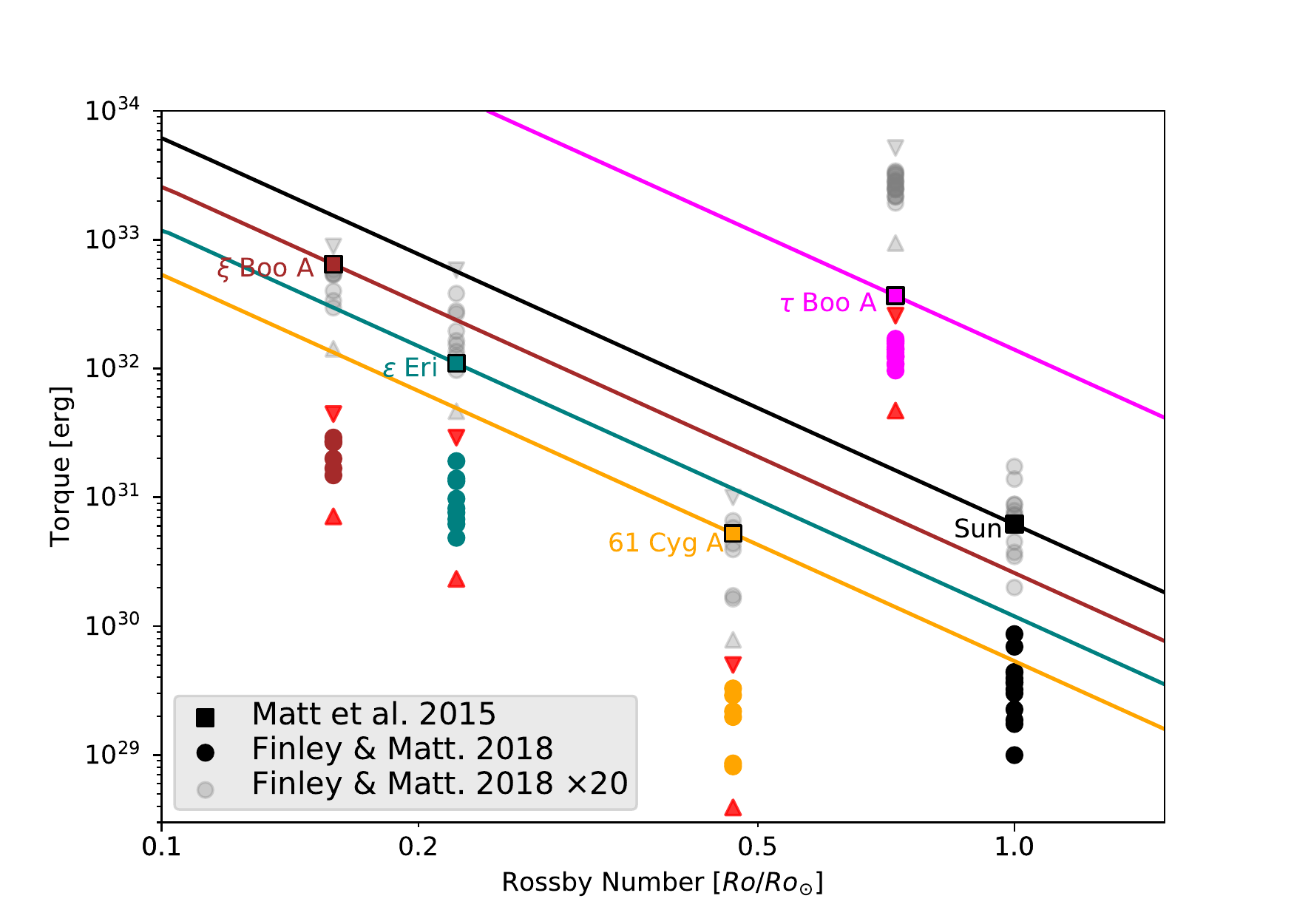}
    \caption{Angular momentum loss rate versus Rossby number. Solid lines represent the \citetalias{matt2015mass} models for each star in our sample, over a range of Rossby numbers. Colored squares indicate the predicted value given our calculated Rossby numbers in Table \ref{StellarParameters}. The torques computed from ZDI epochs and the \citetalias{finley2018dipquadoct} braking law are shown with colored circles. The range of observational uncertainty in the prediction of \citetalias{finley2018dipquadoct} is shown with red limiting triangles. In all cases, the modelled torques using \citetalias{finley2018dipquadoct} are lower when compared to those from \citetalias{matt2015mass}. Multiplication of the \citetalias{finley2018dipquadoct} torques by a factor of 20, shown in grey, roughly brings the models towards agreement.}
    \label{torque_rossby}
\end{figure}

\subsection{Comparison of Torques}
In Figure \ref{torque_rossby} the predictions of \citetalias{matt2015mass} for each star are shown with a range of Rossby numbers using solid lines. We indicate the torque for each star in this model, at its Rossby number from Table \ref{StellarParameters}, with colored squares. The torques using \citetalias{finley2018dipquadoct} and the multiple ZDI epochs are shown with corresponding colored circles. {As with Figure \ref{RA_figure}, typical uncertainties in observed rotation rates ($\pm10\%$), mass loss rates ($\pm10\%$), and field strengths ($\pm1$G) of each star lead to error in the prediction of equations (\ref{torque})-(\ref{DQO_law}). The range of possible torques for each star, given these uncertainties, is indicated with red limits. While these uncertainties are significant, they are not large enough to affect any of our conclusions.} For the dipole-dominated stars, the \citetalias{finley2018dipquadoct} torques appear systematically lower than those expected from \citetalias{matt2015mass}, roughly by a factor of 10-30. Grey points show the result of multiplying all the \citetalias{finley2018dipquadoct} torques by a factor of 20, which brings all of the dipole dominated stars into agreement.

$\tau$ Boo A however, requires a much smaller factor of $\sim3$ to bring the two torques into agreement. Why the torques for $\tau$ Boo A are in better agreement than the other stars is unknown. However, it is worth noting that the mass loss rate for this star has not been measured. {Instead, we used the average mass loss rate from \cite{nicholson2016temporal}, which is directly dependent on their choice of base wind density and temperature. Given that these quantities are not well constrained by observations, the mass loss rates obtained from these simulations are effectively (although indirectly) assumed a-priori. The same is true for all such models.} If the true mass loss rate is smaller than the value used here, the difference between torques may increase, such that we may find a truly systematic value between the two methods for all of the sample stars. If the mass loss rate of $\tau$ Boo A were smaller, its torque may also become dipole dominated like the rest of the sample.

\section{Discussion}
\subsection{Systematic Differences Between the \citetalias{finley2018dipquadoct} and \citetalias{matt2015mass} Torques}
For all the stars in our sample, the torques from \citetalias{finley2018dipquadoct} systematically predicts lower angular momentum loss rates when compared with the rotational evolution torques from \citetalias{matt2015mass}.  In \citetalias{finley2018effect}, this was also the case and we suggested that one possible solution was if the Sun is in a low torque state at present. Since all five stars here are low, it seems unlikely that they would all be in a low state, so a different explanation should be explored.

A systematic difference between the \citetalias{finley2018dipquadoct} and \citetalias{matt2015mass} torques suggest there should be sources of under-estimation in either the MHD modelling, the rotation-evolution models, or observed properties of these stars. \citetalias{finley2018effect} showed that, for the Sun, using the surface field strength leads to a lower torque estimate compared with estimates based on the open magnetic flux, by a factor of $\sim7$. The reason(s) for this remains unclear, perhaps owing to under-estimation of field strengths in magnetograms or if the coronal magnetic field becomes open much closer to the solar surface. {Under-prediction of the open magnetic flux will artificially reduce the braking torque, given the strong correlation shown by \cite{reville2015effect}.}

There are likely also systematics in the magnetic field strengths obtained from ZDI. It is well known that ZDI does not reconstruct all of the photospheric magnetic field due to flux cancellation effects \citep[][See et al., in prep a]{Reiners2009, lehmann2018}. Recently, \cite{2018arXiv181103703L} showed that ZDI sometimes underestimates the field strengths of the large-scale field components, i.e. the dipole, quadrupole and octupole, by a factor of a few. Consequently, the spin-down torques will also be underestimated (also, see discussion by See et al., in prep b). {Additionally, the method used to calculate $\mathcal{R}_{dip}$, $\mathcal{R}_{quad}$, and $\mathcal{R}_{oct}$ from the results of ZDI may lead to under-estimation in the strength of the magnetic field. Given the inherent non-axisymmetry of the ZDI fields, the values we calculate simply approximate the relative strengths of each component. Typically, the polar field values required for the equation (\ref{DQO_law}) will be larger than the global average field strength used in this work, however the effect this has is not large enough to modify our conclusions.}

To increase the \citetalias{finley2018dipquadoct} torques by a factor of 20, for example, would require $\sim 4\times$ greater average Alfv\'en radii, or $\sim26\times$ stronger dipole field strengths than observed. Based on this, it is not clear if this discrepancy can be explained with our current knowledge. Perhaps a combination of wind energetics, as discussed in \citetalias{finley2018effect} for the open flux problem, and the systematics of ZDI might be able to explain the under-prediction of the \citetalias{finley2018dipquadoct} torques versus those of \citetalias{matt2015mass}.

\subsection{The Impact of Magnetic Variability on Dynamical Torque Estimates}
During each sequence of ZDI observations our sample stars experience variability in their global magnetic field strength and topology. In Figure \ref{RA_figure} the predicted average Alfv\'en radii for each ZDI epoch are plotted with a symbol that represents the governing topology in equation (\ref{DQO_law}). In the majority of cases, despite strengthening of the multipolar components, the dipole component governs the location of the torque-averaged Alfv\'en radius.

Similarly, See et al. (in prep b) show for a large range of stars observed with ZDI that equation (\ref{DQO_law}) predicts angular momentum loss rates are dominated by the dipolar component. However for sufficiently high mass loss rates and weak dipolar fields, as seen in this work with $\tau$ Boo A, some stars can have multipolar dominated wind braking. These stars possess low wind magnetisations and so have small average Alfv\'en radii. Note that, if the field strengths are underestimated, as discussed in Section 5.1, even $\tau$ Boo A could then be dipole dominated.

In general, the extrema of the torques from our ZDI stars is $0.5-1.9$ times the average torque, $\langle\tau_{FM18}\rangle$. Using the sub-sampled solar epochs we find the maximum torque to be $2.3 \langle\tau_{FM18}\rangle$. If instead we consider the complete dataset from \citetalias{finley2018effect}, we find the maximum torque is $2.5 \langle\tau_{FM18}\rangle$, slightly larger than the sub-sampled value. Similarly, for other stars, we expect that the true amplitude of variability can be larger than represented by the sparse sampling. The next largest amplitude of variation is found for $\epsilon$ Eri, where in the ZDI epoch of 2013.75 the maximum torque is $1.9 \langle\tau_{FM18}\rangle$. The smallest amplitude of torque variability belongs to $\tau$ Boo A, which has a minimum torque of $0.7 \langle\tau_{FM18}\rangle$,  and a maximum torque of $1.3 \langle\tau_{FM18}\rangle$.

We find results gained by sub-sampling the solar dataset produce average torques which are dependent on the selected magnetogram epochs. For example, by changing the length of the available dataset and selecting a different set of 13 epochs, we can find average torques of $0.3-0.4\times 10^{30}$erg, due to preferentially selecting epochs from cycle 24 or 23 respectively (with 23 being stronger than 24). Equally, reducing the number of epochs used in the dataset from 13 to 6 can change the average torque to a similar degree, but also generally decreases the maximum torque to values comparable to those of the ZDI stars ($\sim2 \langle\tau_{FM18}\rangle$). Reducing the number of epochs further can lead to extreme values in the average torques from $0.1-0.8\times 10^{30}$erg, due to short-term variability in the dataset.

Estimates like this for the Sun hint at how a restricted dataset may bias the time-varying torque estimates for other stars. {Based on the results from this work, it appears that stellar wind variability has a much smaller effect than is required to remedy the discrepancy between stellar wind torques and their long-time rotation evolution counterparts. However, variability does have the ability to confuse the issue and should be accounted for in future works.}

\subsection{Establishing the Timescales of Variability}
In this work we are able to calculate the time-varying torque for four stars with a cadence of $\sim1-2$ years, and over a period of nearly decade. The variation of the torque, due to magnetic variability, can be thought of as an uncertainty in estimating the current average torque for a given star based on a single observation. In \citetalias{finley2018effect}, the variability of the solar wind was examined on a much shorter ($\sim27$ day) cadence over ~2 decades, so that we were able to more continuously estimate the torque. Even so, variability in the solar wind is observed on still shorter timescales. These day to day, and hour to hour, variations in the solar wind are averaged in our calculations in \citetalias{finley2018effect}, in order to better represent the global wind when using observations from a single in-situ location. The impact such fluctuations have on the 27 day torque averages remains an open question.

On timescales of centuries to millennia (still shorter than the braking timescale), there is also evidence for further magnetic variability. For the Sun, indirect methods of detecting this variability, such as examining the concentration of cosmogenic radionuclides (\ce{^{14}C}, \ce{^{10}Be}, etc) in tree trunks or polar ice cores, have been successful at recovering changes in the magnetic field over the last millennia \citep{wusolar}. For other stars, we are unable to examine the evolution of their magnetism for longer than current observations allow. However, the observed spread of magnetic activity indicators \citep[e.g., X-rays;][]{wright2011stellar} around their secular trends, could be caused by variability (as opposed to true differences in stars’ average properties). It is still not clear how such long-term variability may skew our current evaluation of stellar braking torques.

\section{Conclusion}
In this paper we have quantified the effect of observed magnetic variability on the predicted angular momentum loss rates for four Sun-like stars. Our sample stars have all been repeatedly observed with Zeeman-Doppler imaging, which provides information on the topology of the magnetic field. This information is then combined with estimates of their mass loss rates from studies of astrospheric Lyman-$\alpha$, and a relationship for the stellar wind braking given by \citetalias{finley2018dipquadoct}. We compare these time-varying estimates of the angular momentum loss rate to the long-time average value predicted by \citetalias{matt2015mass}, a rotational evolution model.

We find that, similarly to what was found for the Sun in \citetalias{finley2018effect}, the angular momentum loss rates predicted vary significantly (roughly $0.5-1.5$ times their average values), such that torques calculated using single observational epochs can differ from the decadal average torque on the star. This represents an uncertainty when calculating torques for stars with single epochs of observation.

Our calculated angular momentum loss rates based on \citetalias{finley2018dipquadoct} are found to be systematically lower than the long-time average torques required by \citetalias{matt2015mass}. We do not know the origin of this discrepancy, but it could be due (at least in part) to the open flux problem, whereby wind models currently under-predict the observed open magnetic flux for the Sun, problems with observed parameters, such as the potential systematic effects from the ZDI technique in recovering the correct field strengths \citep{2018arXiv181103703L}, problems with rotation-evolution models, or longer-term variability in the torque. Such longer term variability has the potential to affect our predictions for the long-time ($\sim10-100$Myr) average torque required by rotation evolution models.



\acknowledgments
{The authors thank the anonymous referee for their constructive report that helped to improve this work.}
We are also thankful for the ongoing efforts of the ZDI community for making work like this possible.
In particular, we thank Rim Fares and Matthew Mengel for providing the data required to compute the magnetic properties of $\tau$ Boo A.
We thank the SDO/HMI, and SOHO/MDI consortia for providing magnetograms of the Sun.
We thank the ACE/MAG, and ACE/SWEPAM instrument teams, along with the ACE Science Center, for providing in-situ plasma and magnetic field measurements of the solar wind.
This project has received funding from the European Research Council (ERC) under the European Union’s Horizon 2020 research and innovation programme (grant agreement No 682393 AWESoMeStars).
Figures within this work are produced using the python package matplotlib \citep{hunter2007matplotlib}.





\appendix
\section{A - Sun-as-a-Star Data}

Table \ref{Parameters_Solar} displays the selected magnetogram observations from SOHO/MDI and SDO/HMI used in Figure \ref{sun_figure}, and the results of the angular momentum loss calculation using both the formula from \citetalias{finley2018dipquadoct} and \citetalias{matt2015mass}, where symbols have the same meaning as in Table \ref{Parameters}.
\begin{table*}[htbp]
  \centering
\caption{Solar Magnetic Properties and Angular Momentum Loss Results}
\label{Parameters_Solar}
\center
\setlength{\tabcolsep}{1pt}
    \begin{tabular}{c|ccccc|cccc}
    \hline\hline
Star	&	Magnetogram	&	$B_{*}$	&	$\mathcal{R}_{dip}$	&	$\mathcal{R}_{quad}$	&	$\mathcal{R}_{oct}$	&	$\langle R_A \rangle/R_*$	&	$\tau_{FM18} $	&	$\tau_{M15} $	&	$\tau_{M15}$		\\
Name	&	Epoch (Instrument)	&	 (G)	&	$\equiv B_{dip}/B_{*}$	&	$\equiv B_{quad}/B_{*}$	&	$\equiv B_{oct}/B_{*}$	&		&	 ($\times 10^{30}$erg)	&	 ($\times 10^{30}$erg)	&	$\langle \tau_{FM18}\rangle$		\\	\hline
Sun	&	1996.76(MDI)	&	8.0	&	0.38	&	0.11	&	0.51	&	5.9	&	0.87	&	6.20	&	16.55		\\
	&	1998.49(MDI)	&	7.5	&	0.37	&	0.18	&	0.45	&	6.0	&	0.69	&		&			\\
	&	2000.65(MDI)	&	5.7	&	0.22	&	0.16	&	0.62	&	4.3	&	0.30	&		&			\\
	&	2002.37(MDI)	&	8.1	&	0.21	&	0.32	&	0.47	&	5.0	&	0.39	&		&			\\
	&	2004.16(MDI)	&	6.6	&	0.27	&	0.06	&	0.67	&	5.7	&	0.32	&		&			\\
	&	2005.88(MDI)	&	6.1	&	0.32	&	0.15	&	0.53	&	5.3	&	0.44	&		&			\\
	&	2007.59(MDI)	&	5.2	&	0.34	&	0.09	&	0.57	&	5.3	&	0.37	&		&			\\
	&	2009.31(MDI)	&	3.9	&	0.38	&	0.06	&	0.56	&	5.7	&	0.23	&		&			\\
	&	2011.18(HMI)	&	3.1	&	0.30	&	0.33	&	0.37	&	4.3	&	0.17	&		&			\\
	&	2012.89(HMI)	&	2.1	&	0.23	&	0.30	&	0.47	&	3.2	&	0.10	&		&			\\
	&	2014.61(HMI)	&	4.1	&	0.21	&	0.50	&	0.29	&	4.1	&	0.19	&		&			\\
	&	2016.33(HMI)	&	5.7	&	0.31	&	0.29	&	0.40	&	5.2	&	0.36	&		&			\\
	&	2018.12(HMI)	&	5.2	&	0.38	&	0.07	&	0.55	&	5.4	&	0.44	&		&			\\
	\hline
  \end{tabular}
\end{table*}

\section{B - Alternative View of $\tau$ Bootis A Data}
Here we show the result of phase-folding the data from Figure \ref{tauBoo_figure}. $\tau$ Boo A is estimated to have a short magnetic cycle period of around 240 days which is in-phase with its 120 day chromospheric activity cycle. We phase-fold the data for $\tau$ Boo A on the timescale of its chromospheric cycle, rather than its magnetic cycle, as our predictions do not consider the polarity of the magnetic field. Given cycle to cycle variation in length and strength, fitting a simple sinusoid does not well-fit all of the magnetic variation.
\begin{figure}
   \centering
    \includegraphics[trim=0cm 0cm 0cm 0cm, width=0.5\textwidth]{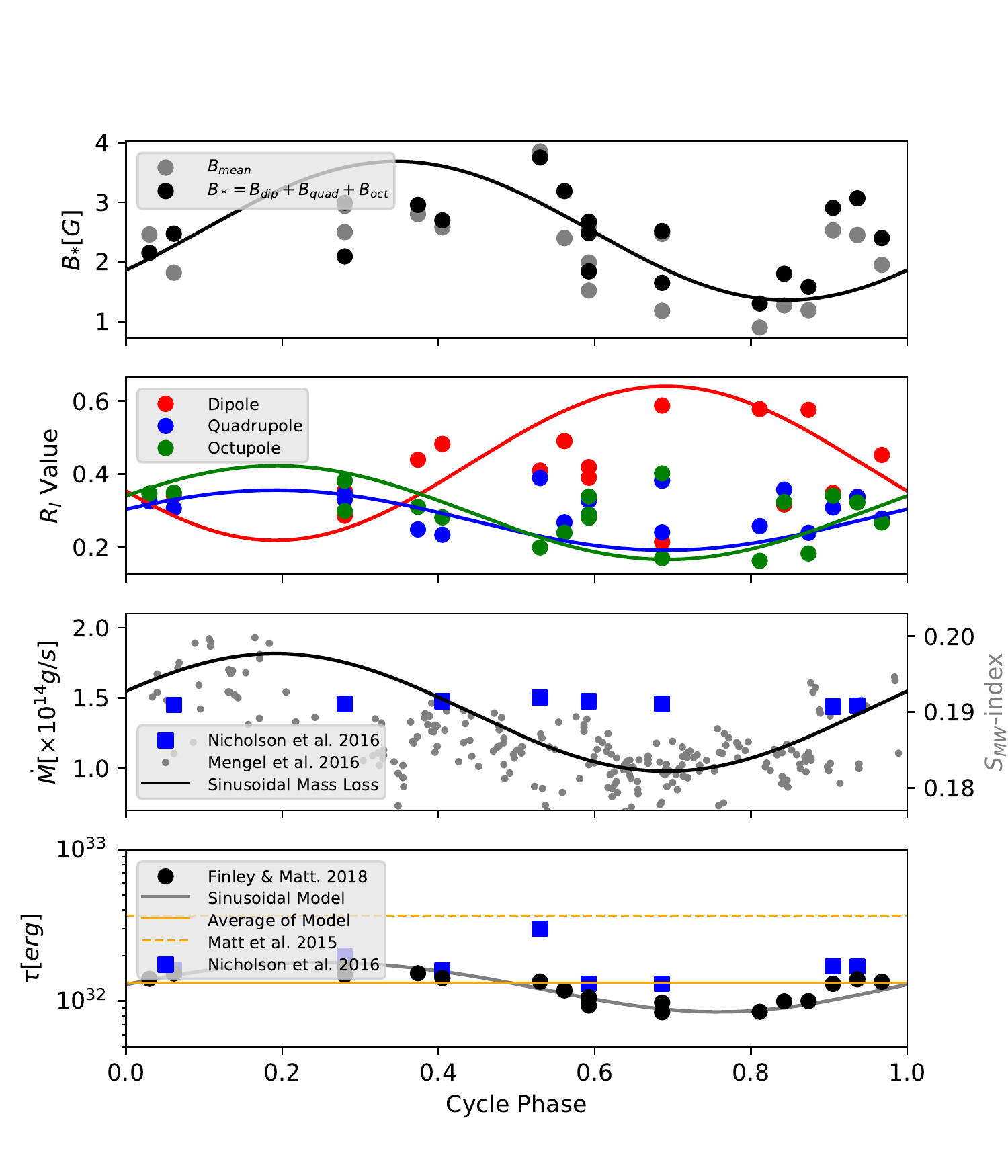}
     \caption{Same as Figure \ref{tauBoo_figure}, with the data phase folded into the 120 day chromospheric cycle.}
     \label{tauBoo_figure2}
\end{figure}



\newpage
\bibliographystyle{yahapj}
\bibliography{CyclesPaper}




\end{document}